\documentclass[fleqn,usenatbib]{mnras}

\usepackage{newtxtext,newtxmath}

\usepackage[T1]{fontenc}
\usepackage{tabularx} 

\usepackage{graphicx}	
\usepackage{amsmath}	

\title[UDGs in Abell~2744 observed by JWST]{Near-infrared characterization of ultra-diffuse galaxies in Abell~2744 by JWST/NIRISS imaging}

\author[Ikeda et al.]{Ryota Ikeda,$^{1,2}$\thanks{E-mail: ryota.ikeda@grad.nao.ac.jp}
Takahiro Morishita,$^{3}$
Takafumi Tsukui, $^{4,5}$
Benedetta Vulcani,$^{6}$
Michele Trenti,$^{5,7}$
\newauthor 
Benjamin Metha,$^{5,7,8}$
Ana Acebron,$^{9,10}$
Pietro Bergamini,$^{9,11}$
Claudio Grillo,$^{9,10}$
Daisuke Iono,$^{1,2}$
\newauthor 
Amata Mercurio,$^{12}$
Piero Rosati,$^{11,13}$ and Eros Vanzella$^{11}$ \\
$^{1}$Department of Astronomy, School of Science, SOKENDAI (The Graduate University for Advanced Studies), 2-21-1 Osawa, Mitaka, Tokyo 181-8588, Japan\\
$^{2}$National Astronomical Observatory of Japan, 2-21-1 Osawa, Mitaka, Tokyo 181-8588, Japan\\
$^{3}$IPAC, California Institute of Technology, MC 314-6, 1200 E. California Boulevard, Pasadena, CA 91125, USA\\
$^{4}$Research School of Astronomy and Astrophysics, Australian National University, Cotter Road, Weston Creek, ACT 2611, Australia\\
$^{5}$ARC Centre of Excellence for All Sky Astrophysics in 3 Dimensions (ASTRO 3D)\\
$^{6}$INAF Osservatorio Astronomico di Padova, vicolo Osservatorio 5, 35122 Padova, Italy\\
$^{7}$School of Physics, University of Melbourne, Parkville 3010, VIC, Australia\\
$^{8}$Department of Physics and Astronomy, University of California, Los Angeles, 430 Portola Plaza, Los Angeles, CA 90095, USA\\
$^{9}$Dipartimento di Fisica, Università degli Studi di Milano, Via Celoria 16, I-20133 Milano, Italy\\
$^{10}$INAF - IASF Milano, via A. Corti 12, I-20133 Milano, Italy\\
$^{11}$INAF - OAS, Osservatorio di Astrofisica e Scienza dello Spazio di Bologna, via Gobetti 93/3, I-40129 Bologna, Italy\\
$^{12}$INAF -- Osservatorio Astronomico di Capodimonte, Via Moiariello 16, I-80131 Napoli, Italy\\
$^{13}$Dipartimento di Fisica e Scienze della Terra, Università degli Studi di Ferrara, Via Saragat 1, I-44122 Ferrara, Italy\\
}

\date{Accepted XXX. Received YYY; in original form ZZZ}

\pubyear{2022}

\begin{document}
\label{firstpage}
\pagerange{\pageref{firstpage}--\pageref{lastpage}}
\maketitle

\begin{abstract}
We present a search and characterization of ultra-diffuse galaxies (UDGs) in the Frontier Fields cluster Abell~2744 at $z=0.308$. We use JWST/NIRISS F200W observations, acquired as part of the GLASS-JWST Early Release Science Program, aiming to characterize morphologies of cluster UDGs and their diffuse stellar components.
A total number of 22 UDGs are identified by our selection criteria using morphological parameters, down to stellar mass of $\sim10^{7}M_{\odot}$. The selected UDGs are systematically larger in effective radius in F200W than in HST/ACS F814W images, which implies that some of them would not have been identified as UDGs when selected at rest-frame optical wavelengths. In fact, we find that about one third of the UDGs were not previously identified based on the F814W data. 
We observe a flat distribution of the UDGs in the stellar mass-size plane, similar to what is found for cluster quiescent galaxies at comparable mass.  
Our pilot study using the new JWST F200W filter showcases the efficiency of searching UDGs at cosmological distances, with 1/30 of the exposure time of the previous deep observing campaign with HST. Further studies with JWST focusing on spatially-resolved properties of individual sources will provide insight into their origin.
\end{abstract}




\section{Introduction}
The nature and the origin of ultra-diffuse galaxies (UDGs) have been extensively investigated since its large population has been discovered in nearby galaxy clusters (e.g., \citealp{2015ApJ...798L..45V}; \citealp{2015ApJ...807L...2K}; \citealp{2015ApJ...809L..21M}; \citealp{2015ApJ...813L..15M}; \citealp{2016ApJS..225...11Y}; \citealp{2017A&A...608A.142V}; \citealp{2022A&A...665A.105L}). UDGs were originally defined as galaxies with low central surface brightness ($24\leq\mu(g,0)\leq26$~mag~arcsec$^{-2}$) and extended sizes ($R_{e}>1.5$~kpc) in $g$-band  \citep{2015ApJ...798L..45V}. {Alternatively, UDGs have been defined with mean effective surface brightness \footnote{An effective surface brightness ($\langle\mu_{e}\rangle$) denotes the averaged value of surface brightness ($\mu$) within the effective radius ($R_{e}$). We adopt this definition throughout the paper.}($24.0\leq\langle\mu_{e}(r)\rangle\leq26.5$~mag~arcsec$^{-2}$) and extended size ($R_{e}\geq1.5$~kpc) in the $r$-band \citep{2016A&A...590A..20V}.} Some UDGs weres identified as an extreme case of low-surface brightness galaxies (LSBGs; \citealp{1997ARA&A..35..267I}) {without apparent disks,} over several decades (e.g., \citealp{1984AJ.....89..919S}; \citealp{1987AJ.....94...23B}; \citealp{1988ApJ...330..634I}; see \citealp{2016ApJS..225...11Y} for the compilation). While their sizes are comparable to the ones of Milky-Way like galaxies, the stellar masses of UDGs are typically 2--3 orders of magnitudes smaller (e.g., \citealp{2018MNRAS.473.3747S}), which makes the population unique and puzzling.

Importantly, recent studies uncovered that UDGs comprise a large population in massive clusters, in which their number is comparable to bright galaxies down to $M_{r} < M_{r}^{*} + 2.5$ \citep{2017A&A...607A..79V}. However, the origin and formation pathways of UDGs are still unknown. The possible formation mechanisms of UDGs can be roughly classified into two types; external and internal processes. 
As an example of external processes, \cite{2019MNRAS.485..382C} show that {in their modeling using cosmological dark matter-only simulation,} a mass-loss due to the tidal interaction, and expansion of stellar component as a consequence of tidal heating can reproduce observational properties of cluster UDG population, such as stellar masses, size distributions, and abundances (see also \citealp{2019MNRAS.487.5272J}; \citealp{2020MNRAS.494.1848S}; \citealp{2020MNRAS.497.2786T}). Internal processes such as the retention of high spin halos (\citealp{2016MNRAS.459L..51A}; \citealp{2017MNRAS.470.4231R}) and strong stellar feedback (\citealp{2017MNRAS.466L...1D}; \citealp{2018MNRAS.478..906C}) might also explain the properties of UDGs. However, conflicts with observations have recently been claimed for the feedback model by \cite{2022ApJ...941...11K}. They discuss that relatively extended SFR distribution compared to low-mass dwarf galaxies and the lack of current vigorous star formation activity in the H{\sc i}-selected UDGs are in conflict with the picture in which feedback-driven expansion is crucial.

Whether high-density environment is crucial for the formation of UDGs is an important open question. On the one hand, it is likely that the abundance of UDGs as a function of halo mass is sub-linear, implying that there is no major environmental effect on the formation and survival of UDGs (\citealp{2016A&A...590A..20V}; \citealp{2017MNRAS.468.4039R};
\citealp{2017ApJ...844..157L};
\citealp{2018MNRAS.481.4381M}; \citealp{2020ApJ...894...75L}; 
\citealp{2022arXiv221014994L}; \citealp{2023MNRAS.519..884K}, but see \citealp{2017A&A...607A..79V}; \citealp{2019ApJ...887...92J}). On the other hand, evidence of environmental dependencies of UDGs have also been reported. Compared to field counterparts, cluster UDGs tend to be red \citep[e.g.,][]{2019MNRAS.488.2143P}, old (e.g., \citealp{2020ApJS..247...46B}; \citealp{2021ApJ...923..257K}; \citealp{2022MNRAS.517.2231B}), and {globular cluster-rich (e.g., \citealp{2020MNRAS.492.4874F}; \citealp{2020ApJ...902...45S}).} These trends suggest that it is likely - or as a minimum possible -  that the environment influences the formation paths of UDGs. Therefore, in concordance with a variety of theoretical predictions, it has recently been suggested that the UDG population consists of a mixture of different origins (e.g., \citealp{2019MNRAS.490.5182L}; \citealp{2020MNRAS.494.1848S}; \citealp{2020ApJ...894...75L};  \citealp{2022MNRAS.517.2231B}), or maybe formed by a combination of internal and external processes \citep{2019MNRAS.485..796M}.

To assess the origin of UDGs, it would be of interest to identify such populations in the distant universe and to investigate their properties.
As possible progenitors, those distant counterparts may provide us a new insight into their origin as they might have fallen into the gravitational potential of a cluster halo more recently. 
These investigations have been pioneered on LSBGs, as likely progenitors of UDGs, up to $z\sim1.2$, using the deep image of stacked HST data \citep{2021A&A...646L..12B}. Nonetheless, these attempts require deep images, predominantly due to the cosmological dimming of surface brightness. In addition, since UDGs in the nearby universe have been conventionally observed and studied at optical wavelengths, deep near-infrared (NIR) imaging of progenitor UDGs would be essential for a fair comparison with local counterparts.

In this context, the sensitivity and spatial resolution of JWST in NIR wavelengths is expected to be a powerful probe for a detailed study of diffuse galaxies at cosmological distances. In addition to its sensitivity to faint light enabled by the large mirror, NIR imaging serves as an effective tracer of stellar mass distribution, given the stellar mass-to-light ratio is expected to be close to unity at these wavelengths \citep[e.g.,][]{2001ApJ...550..212B}. \cite{2023arXiv230304726C} report a sample of possible progenitors of local cluster UDGs in the El Gordo cluster at $z=0.87$, identified from the stacked image of eight JWST/NIRCam filters.

In this paper, we report on the identification and characterization of UDGs in Abell~2744 ($z=0.308$) selected from a rest-frame NIR wavelength using the JWST/NIRISS F200W imaging taken by the GLASS Early Release Science program \citep{2022ApJ...935..110T}. UDGs in Abell~2744 have been searched before with rest-frame optical wavelengths, using the Hubble Space Telescope (HST) imaging (\citealp{2017ApJ...839L..17J}; \citealp{2017ApJ...844..157L}). This paper aims to address how the sample of UDGs differ from the optically-selected UDGs when we select them in NIR wavelengths, and to characterize the morphologies and structural parameters in different wavelengths. We first give a brief description of the JWST data in Section~\ref{sec:Sec2}. In Section~\ref{sec:Sec3}, we describe the photometric analyses. We present the sample selection of UDGs and their characterization in Section~\ref{sec:Sec4}. We discuss the comparison of UDGs selected in different wavelengths 
and the stellar mass-size relation in Section~\ref{sec:Sec5}. We summarize this study in Section~\ref{sec:Sec6}. 

Throughout this paper, we assume a flat $\Lambda$CDM cosmology and adopt the cosmological parameters of $H_{0}=70 \ {\rm km \ s^{-1} \ Mpc^{-1}}$, $\Omega_{M}=0.3$, and $\Omega_{\Lambda}=0.7$. A redshift of $z=0.308$ corresponds to a cosmic age of 9.97\,Gyr and gives a projected physical scale of 4.536\,kpc/$''$. 
{We refer to the term "size" as the circularized effective radius, that is the product of the major-axis effective radius from the best-fit S\'{e}rsic model and $\sqrt{q}$, where $q$ is the minor-to-major axis ratio.}

\section{Data} \label{sec:Sec2}

We use JWST/NIRISS imaging data taken as part of the GLASS-JWST early release science (ERS) program (\citealp{2022ApJ...935..110T}). NIRISS observations of the Abell~2744 cluster field ($2\farcm2\times2\farcm2$) were conducted in three filters of F115W, F150W, and F200W. The NIRISS field is entirely covered with several HST programs (\citealp{2015ApJ...812..114T}; \citealp{2017ApJ...837...97L}; \citealp{2020ApJS..247...64S}). The NIRISS observations were executed on 2022 June 28-29 with an exposure time of 2830s for each filter, reaching a depth of $\sim$28.6-28.9 AB mag at $5\sigma$.

We use the reduced images, including archival HST data, presented in \cite{2022ApJ...938L..13R}. Briefly, all imaging data in the FoV were consistently reduced by using {\tt grizli} software \citep{2021zndo...1146904B}. {\tt grizli} includes extra processes for NIRISS data, such as 1/f-noise subtraction and masking of optical ghosts. We resampled the images in a common pixel scale of $0\farcs066$. In the last step, we create an infrared-stack image, by combining F125W, F150W, F160W, and F200W images (Section~\ref{subsec:Sec3.1}).

In this study, we choose the F200W filter for selecting UDGs. As the cluster UDGs in the nearby universe are known to have an old ($\gtrsim7$~Gyr) stellar population (e.g., \citealp{2018ApJ...858...29P}; \citealp{2018MNRAS.478.2034R}; 
\citealp{2018MNRAS.479.4891F}; \citealp{2018ApJ...859...37G};
\citealp{2020ApJS..247...46B}; \citealp{2021ApJ...923..257K}; \citealp{2022MNRAS.517.2231B}), it can be expected that cluster UDGs become brighter in NIR wavelength. At the redshift of Abell~2744, F200W corresponds to the rest-frame wavelength of $\sim1.6$\,\micron~{($H$-band)}, which is dominated by continuum emission from old stellar populations and thus reflects the stellar mass distribution in the system \citep{2001ApJ...550..212B}. Moreover, the sensitivity of the F200W imaging is the deepest (28.9 AB mag at $5\sigma$) among the three JWST/NIRISS filters available in the field. {To compare with the F200W filter, we also use the HST/ACS F814W filter, which provides the rest-frame wavelength of $\sim0.5$\,\micron~{($V$-band)} for the cluster members.}


\section{Photometric Analysis} \label{sec:Sec3}

\subsection{Source extraction and photometric redshift} \label{subsec:Sec3.1} 

We use {\tt borgpipe} \citep{2021ApJS..253....4M} to build a photometric catalog. {\tt borgpipe} starts with running {\tt SExtractor} \citep{1996A&AS..117..393B} on the PSF-matched images, by using the infrared-stack image created in Section~\ref{sec:Sec2} as the detection image. During the process, noise correction caused by drizzling is also taken into account (\citealp{2011ApJ...727L..39T}; see also \citealp{2018ApJ...867..150M}).  Fluxes of all images are extracted within $r=0\farcs32$ aperture. To compensate the aperture loss, the fluxes are then uniformly multiplied by a scaling factor { defined as} $C = f_{\rm det,AUTO}/f_{\rm det,aper.}$ for each source, where $f_{\rm det,AUTO}$ is FLUX\_AUTO and $f_{\rm det,aper.}$ is aperture flux measured in the detection band. { The corrected flux in filter {\tt x} is thus derived as $f_{\rm x} = C  f_{\rm x, aper.}$.}

A total of 3520 sources are detected in the area covered by the JWST/NIRISS data. On these sources, we run the photometric redshift code {\tt EAzY} \citep{2008ApJ...686.1503B}, with a redshift range of [0.01, 20] with a logarithmically equal step size of $\log (1+z) = 0.01$. We then cross-match with the spectroscopic redshift catalogs (\citealp{2018MNRAS.473..663M}; \citealp{2021A&A...646A..83R}),
and find 401 objects matched within $0\farcs5$.
Hereafter, we use the spectroscopic redshifts from the catalog for those available and the photometric redshifts, $z_{\rm phot}$ ($z_{a}$), for the rest of the sources. {For spectroscopically confirmed galaxies at $z<0.5$, we find a median offset of $|z_{\rm phot}-z_{\rm spec}|/(1+z_{\rm spec})=0.036$, securing the accuracy of $z_{\rm phot}$ estimates.}

\subsection{Stellar mass estimates from SED fit} \label{subsec:Sec3.2}

In order to investigate the distributions of the UDGs in the stellar mass-size plane (Section \ref{subsec:Sec5.3}), we derive the stellar masses from spectral energy distribution (SED) fit using {\tt FAST++} code \footnote{\url{https://github.com/cschreib/fastpp}} \citep{2009ApJ...700..221K}. {For the SED fitting, 11 filters are used in total: HST/ACS F435W, F606W, F814W, HST/WFC3 F336W, F105W, F125W, F140W, F160W, and JWST/NIRISS F115W, F150W, F200W.} We assume a delayed-tau star formation history and adopt the {\tt fsps} stellar population library \citep{2009ApJ...699..486C} generated for the Kroupa initial mass function \citep{2001MNRAS.322..231K}. We adopt the Calzetti dust extinction law \citep{2000ApJ...533..682C}. We fix redshift to the values derived from {\tt EAzY} (Section \ref{subsec:Sec3.1}), and the rest of the parameters are treated as free parameters. 

\section{Identification of Ultra Diffuse Galaxies from Near-infrared wavelengths} \label{sec:Sec4}

We follow a two-step selection \citep{2016A&A...590A..20V}, namely selections using {\tt SExtractor} and {\tt GALFIT} \citep{2002AJ....124..266P} to identify UDGs in Abell~2744. We start with excluding obvious non-UDG candidates by using structural properties, half-light radius and total magnitude, obtained from {\tt SExtractor}. However, since {\tt SExtractor} does not correct for PSF convolution, those measurements should only be considered as rough estimates. Therefore, we augment the analysis through a second step with {\tt GALFIT}, which is computationally heavy but provides accurate estimates from a deconvolved model fitting. Lastly, we carefully inspect all fitting results and exclude objects with obvious data-quality issues, such as those severely affected by neighbors from the final sample.

\subsection{Pre-selection of UDG candidates with SExtractor measurements}
\label{subsec:Sec4.1}

\begin{figure}
    \centering
	\includegraphics[width=0.92\columnwidth]{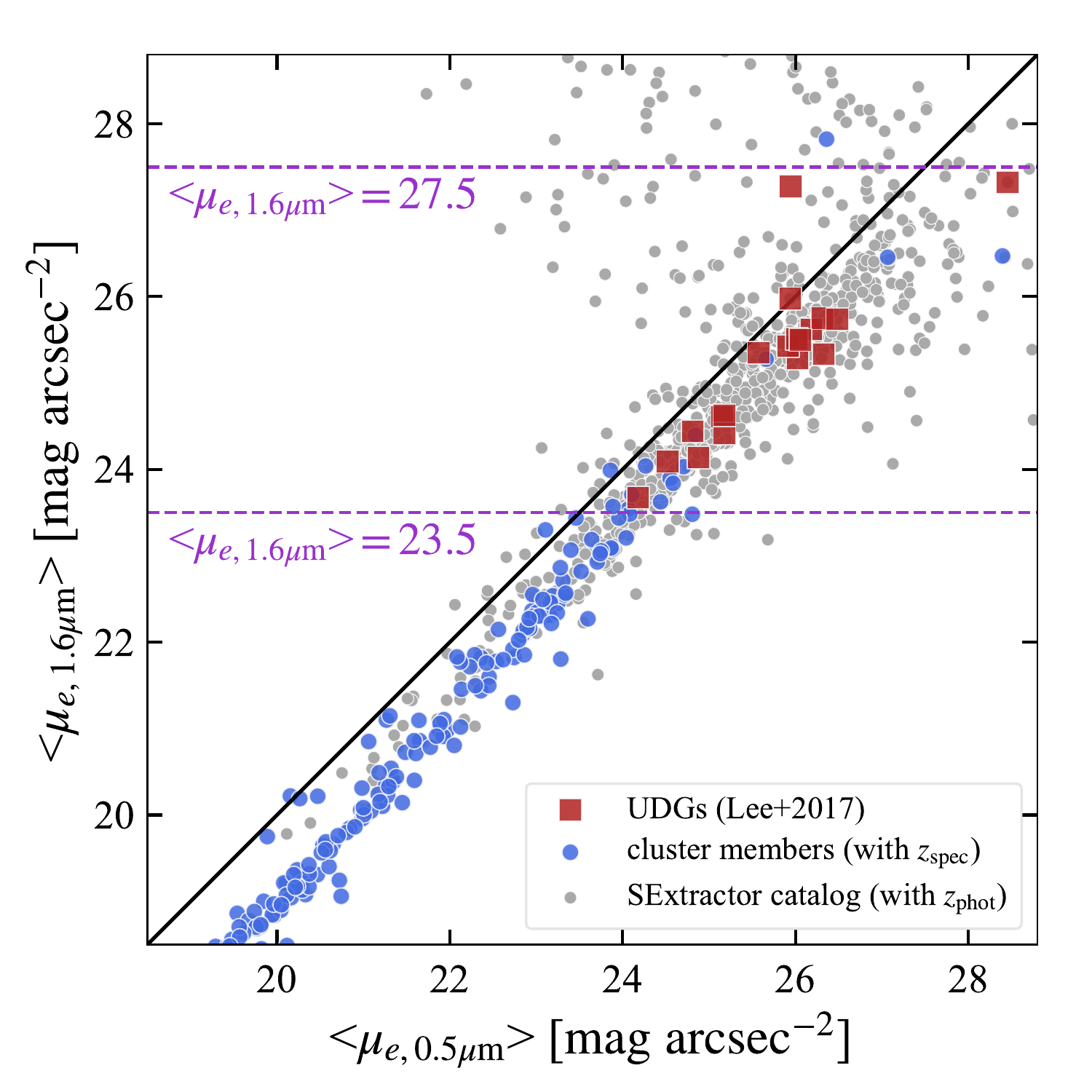}
    \caption{Comparison of surface brightness between HST/ACS F814W and JWST/NIRISS F200W filters. The red symbols are UDGs reported in \citet{2017ApJ...844..157L}. The blue symbols are the compilation of spectroscopically confirmed cluster galaxies (\citealp{2018MNRAS.473..663M}; \citealp{2021A&A...646A..83R}).
    The grey symbols are objects detected in {\tt SExtractor} with $0.108<z_{\mathrm{phot}}<0.508$ (Section~\ref{subsec:Sec3.1}). The surface brightness in the F200W filter is systematically brighter than the F814W filter. We adopt a criterion of ${23.5<\langle\mu_{e,1.6\mu {\rm m}}\rangle<27.5}$~mag~arcsec$^{-2}$ (dashed purple) for reducing the candidates of UDGs.}
    \label{fig:Figure1}
\end{figure}

By using the {\tt SExtractor} output parameters of F200W imaging and redshift information derived from {\tt EAzY} (Section~\ref{subsec:Sec3.1}), we pre-select the sample on which we pursue the {\tt GALFIT} analysis in Section~\ref{subsec:Sec4.2}.
As an initial selection, we adopt FLUX\_RADIUS~$>4.01$ pix (corresponding to 1.2~kpc at $z=0.308$), ELONGATION~$>0.25$, FLAGS~$<4$, and ${23.5<\langle\mu_{e,1.6\mu {\rm m}}\rangle<27.5}~\rm{mag~arcsec^{-2}}$. As we define the final UDGs as galaxies with ${R_{e,1.6\mu {\rm m}}>1.5}$~kpc in the following step, the lower boundary of $1.2$~kpc for FLUX\_RADIUS, representing half-light radius, was chosen on the assumption of 20~\% uncertainties. The selection of FLAGS removes the objects with saturated pixels or with deblending failure, and the truncated objects at the edge of the image. We further request $0.288<z_{\mathrm{spec}}<0.328$ when available, and otherwise $0.108<z_{\mathrm{phot}} (z_{a})<0.508$. We adopt $4\sigma_{\rm clus}$ range for spectroscopic redshift, where $\sigma_{\rm clus}=1497$~km/s is a velocity dispersion of member galaxies of Abell~2744 \citep{2011ApJ...728...27O}.  For photometric redshift selection, we adopt a wider range to roughly exclude the background and nearby sources. {We note that the sources selected through the photometric redshift selection have a median uncertainty of $\Delta z=[-0.179, 0.123]$ in $1\sigma$ confidence level, which is few times smaller than the range we adopted for the sample selection.} 

Figure \ref{fig:Figure1} shows the comparison between the surface brightness of HST/ACS F814W and JWST/NIRISS F200W. We calculate the effective radius by ${\langle\mu_{e}\rangle=m+2.5\log_{10}{(2\pi R_{e}^{2})}}$, following \cite{2005PASA...22..118G}. We find that ${\langle\mu_{e,1.6\mu{\rm m}}\rangle}$ is prone to be brighter than ${\langle\mu_{e,0.5\mu{\rm m}}\rangle}$. We match our {\tt SExtractor} catalog and the UDGs reported in \cite{2017ApJ...844..157L}, which were selected from the HST/ACS F814W imaging, within $0\farcs5$ aperture. A total of 19 UDGs are matched. Based on the distribution of these sources in Figure \ref{fig:Figure1}, we adopt ${\langle\mu_{e,1.6\mu{\rm m}}\rangle}$ of 23.5 and 27.5~mag~arcsec$^{-2}$ as a lower and upper boundaries, where all of the matched UDGs reported in \cite{2017ApJ...844..157L} are included in range. These selections resulted in 225 objects with $z_{\mathrm{phot}}$ and 11 objects with $z_{\mathrm{spec}}$ (236 objects in total).

\subsection{Final selection of UDGs with GALFIT} \label{subsec:Sec4.2}

To identify UDGs and to constrain their structural parameters, we run {\tt GALFIT}, assuming a single S\'{e}rsic model. Although the sky subtraction has already been implemented in the reduced image \citep{2022ApJ...938L..13R}, we include the sky component as the second fitting component to minimize the contamination from the intracluster light \citep{2017ApJ...835..254M}, which is non-negligible especially around the cluster center. For each object, we generate a cutout image of $68\times68$ pixels$^2$ ($\sim20\times20$ kpc$^{2}$ in a physical scale), as well as the root mean square map and segmentation map in the same region. To avoid contamination from the outskirts of nearby objects, we mask nearby sources by using the segmentation map. In the few cases in which the outskirts of nearby objects are blended to the targets, we add another S\'{e}rsic component to diminish their effect. 

A total of 180 out of 236 objects are successfully fitted. For the remaining 56 objects, we find that the error estimates of FLUX\_RADIUS are larger than the fitted value itself, and some parameters failed to converge within the boundary of the fitting. We conclude that the fitting output for these objects is unreliable, and exclude them from the following analysis. These objects comprise a spurious source, and a patchy edge or a companion of bright galaxies. 

We further exclude 27 objects by visual inspection, as either a spurious source or a part of diffuse light of bright extended galaxies are evident, even though the fitting values appear plausible. During this process, we find that in 12 objects, the residual of the best-fit S\'{e}rsic model from the data is prominent due to their disturbed morphology. Three of them are part of a large merging system, which is identical with A2744-DSG-z3, a dusty spiral galaxy at $z=3.059$ reported by \cite{2023ApJ...942L...1W}. They discuss that there is a tentative foreground contamination from A2744-DSG-z3 based on their best-fit photometric redshift of $z\sim0.4$, which explains why this galaxy remained in our list of UDG candidates. To securely evaluate the structural parameters of UDGs, we excluded these 12 objects (10 systems in total) from being classified as UDG candidates.
In summary, after processing the sample selections via {\tt SExtractor} and {\tt GALFIT}, we obtain the structural parameters of UDG candidates for 141 objects.

\begin{figure}
	\includegraphics[width=\columnwidth]{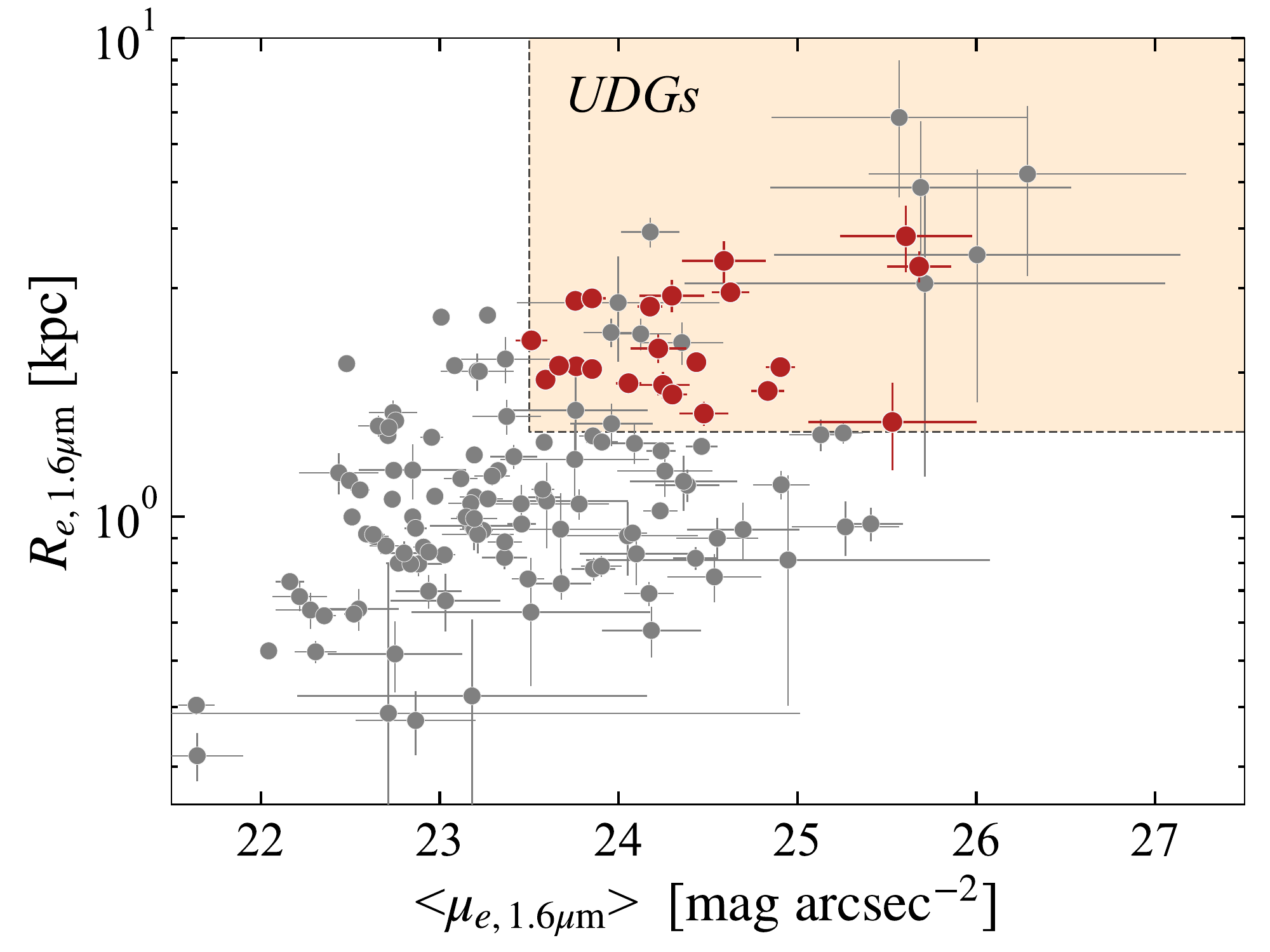}
    \caption{The location of the UDGs (red circles) in the effective surface brightness (${\langle\mu_{e,1.6\mu {\rm m}}\rangle}$) - the circular effective radius (${R_{e,1.6\mu {\rm m}}}$) plane. The highlighted region in beige signifies the criteria of UDGs. The rest of non-UDGs are shown in grey circles. A total number of 34 out of 141 objects fulfill the criteria, and 12 of them are excluded from the final UDG  due to their high S\'{e}rsic index (${n_{1.6\mu{\rm m}}>2.5}$).}
    \label{fig:Figure2}
\end{figure}

\begin{figure}
    \centering
	\includegraphics[width=0.9\columnwidth]{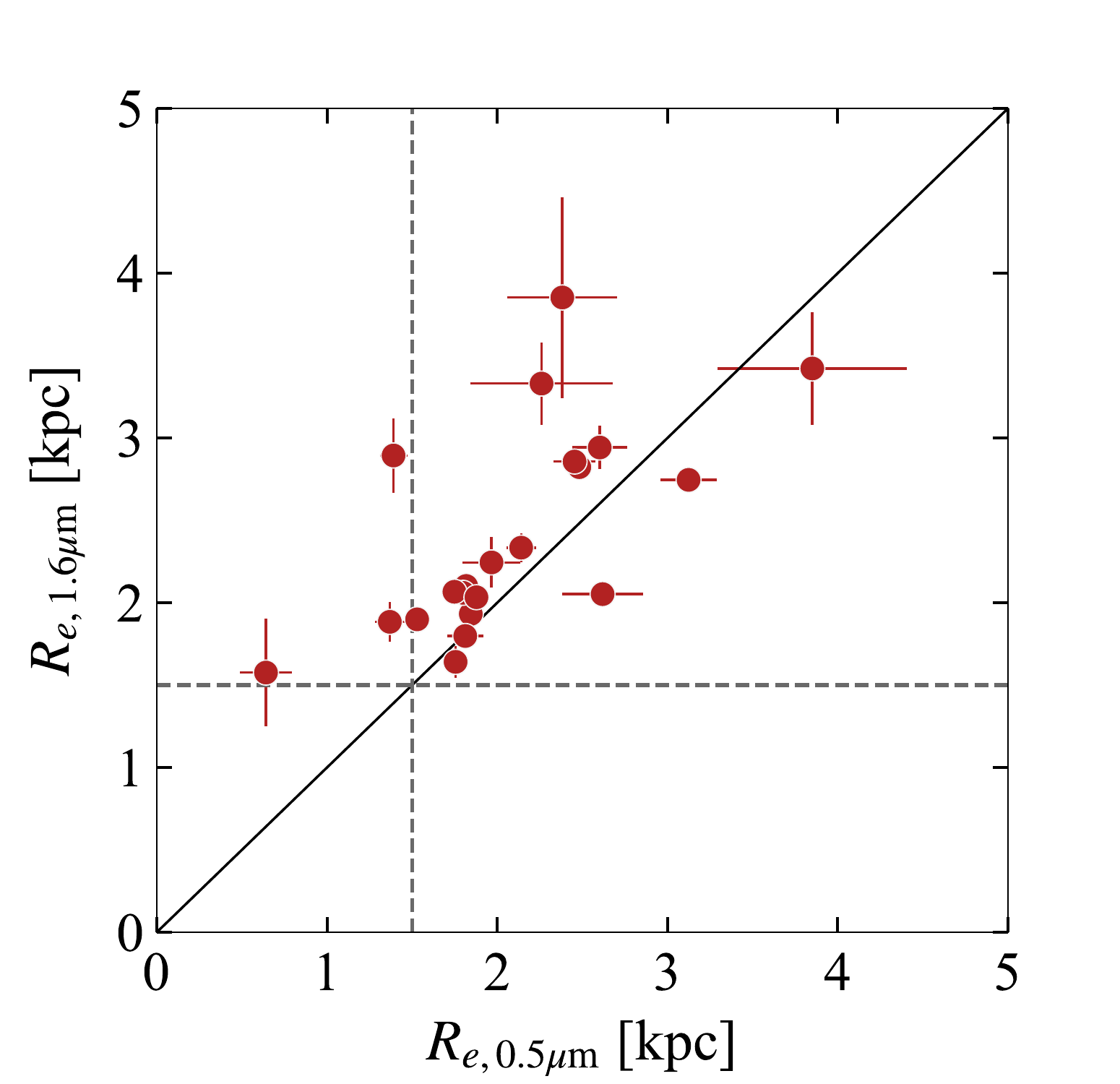}
    \caption{Comparison of the circular effective radius of the UDGs, measured in the F200W and F814W filters. The one-to-one relation is shown in the black solid line. The dashed lines indicate $R_{e}=1.5$ kpc.}
    \label{fig:Figure3}
\end{figure}

\begin{table*}
	\centering
	\fontsize{8pt}{4.3mm}\selectfont
	\caption{The properties of UDGs selected from the rest-frame 1.6~\micron \ wavelength}
	\label{tab:Table1}
	\begin{tabular}{ccccccccccc}
		\hline\hline
		& & & \multicolumn{3}{c}{F200W} & & \multicolumn{3}{c}{F814W} & \\ \cline{4-6} \cline{8-10} 
		ID & R.A. & Decl. &${R_{e,1.6\mu {\rm m}}}$ &${\langle\mu_{e,1.6\mu{\rm m}}\rangle}$ &${n_{1.6\mu{\rm m}}}$ & &${R_{e,0.5\mu {\rm m}}}$ &${\langle\mu_{e,0.5\mu{\rm m}}\rangle}$ &${n_{0.5\mu{\rm m}}}$ &$\log({M_{\star}/M_{\odot})}$  \\
		   & (deg) & (deg) & (kpc) & (mag~arcsec$^{-2}$) & & & (kpc) & (mag~arcsec$^{-2}$) & &  \\
		\hline
1249 & 3.58160 & -30.41289 & $2.94\pm0.13 $ & $24.63\pm0.11 $ & $1.29\pm0.06 $ & & $2.60\pm0.16 $ & $25.18\pm0.14 $ & $1.37\pm0.09 $ & $7.78 _{- 0.01 }^{+ 0.04 }$ \\
1675 & 3.59833 & -30.41287 & $2.05\pm0.07 $ & $24.90\pm0.08 $ & $0.61\pm0.04 $ & & $2.62\pm0.24 $ & $25.73\pm0.22 $ & $1.11\pm0.11 $ & $7.24 _{- 0.03 }^{+ 0.05 }$ \\
1916 & 3.58420 & -30.41006 & $1.90\pm0.06 $ & $24.06\pm0.07 $ & $1.26\pm0.04 $ & & $1.53\pm0.06 $ & $24.28\pm0.08 $ & $1.20\pm0.06 $ & $7.87 _{- 0.03 }^{+ 0.02 }$ \\
2114 & 3.58692 & -30.40987 & $2.24\pm0.15 $ & $24.22\pm0.16 $ & $2.32\pm0.13 $ & & $1.97\pm0.17 $ & $24.63\pm0.20 $ & $2.10\pm0.17 $ & $7.39 _{- 0.05 }^{+ 0.01 }$ \\
2494 & 3.60566 & -30.40536 & $2.82\pm0.05 $ & $23.76\pm0.04 $ & $1.21\pm0.02 $ & & $2.48\pm0.06 $ & $24.24\pm0.06 $ & $1.13\pm0.03 $ & $8.34 _{- 0.02 }^{+ 0.04 }$ \\
2706 & 3.60307 & -30.40571 & $3.33 \pm 0.25$ & $25.68\pm0.18$ & $1.36 \pm 0.10$ & & $2.29 \pm 0.43$ & $25.90 \pm 0.45$ & $0.82 \pm 0.42$ &  $8.26 _{- 0.23 }^{+ 0.40 }$ \\
3073 & 3.61693 & -30.40145 & $2.86\pm0.09 $ & $23.85\pm0.08 $ & $1.73\pm0.05 $ & & $2.45\pm0.12 $ & $24.56\pm0.12 $ & $1.62\pm0.08 $ & $8.32 _{- 0.04 }^{+ 0.05 }$ \\
3396 & 3.59815 & -30.39890 & $3.85\pm0.61 $ & $25.61\pm0.37 $ & $2.15\pm0.22 $ & & $2.38\pm0.32 $ & $25.55\pm0.31 $ & $1.73\pm0.22 $ & $7.88 _{- 0.04 }^{+ 0.05 }$ \\
3626 & 3.60776 & -30.39523 & $2.10\pm0.04 $ & $24.43\pm0.04 $ & $0.55\pm0.02 $ & & $1.82\pm0.05 $ & $24.98\pm0.07 $ & $0.35\pm0.04 $ & $7.65 _{- 0.08 }^{+0.01 }$ \\
3637 & 3.57268 & -30.39634 & $1.93\pm0.03 $ & $23.59\pm0.04 $ & $0.81\pm0.02 $ & & $1.84\pm0.03 $ & $24.03\pm0.04 $ & $0.83\pm0.02 $ & $8.28 _{- 0.03 }^{+ 0.02 }$ \\
3896 & 3.60125 & -30.39521 & $1.88\pm0.12 $ & $24.25\pm0.15 $ & $1.37\pm0.09 $ & & $1.37\pm0.09 $ & $24.48\pm0.15 $ & $1.13\pm0.10 $ & $7.69 _{- 0.02 }^{+ 0.04 }$ \\
4426 & 3.57906 & -30.38928 & $2.06\pm0.07 $ & $23.76\pm0.08 $ & $1.78\pm0.06 $ & & $1.80\pm0.12 $ & $24.28\pm0.15 $ & $1.90\pm0.12 $ & $8.51 _{- 0.04 }^{+ 0.01 }$ \\
4464 & 3.58152 & -30.39016 & $2.33\pm0.09 $ & $23.51\pm0.09 $ & $1.78\pm0.06 $ & & $2.14\pm0.09 $ & $23.91\pm0.09 $ & $1.57\pm0.06 $ & $8.27 _{- 0.02 }^{+ 0.01 }$ \\
4512 & 3.61281 & -30.39049 & $2.07\pm0.05 $ & $23.67\pm0.05 $ & $1.08\pm0.03 $ & & $1.75\pm0.03 $ & $23.74\pm0.05 $ & $1.00\pm0.02 $ & $7.86 _{-0.03 }^{+ 0.16 }$ \\
4692 & 3.57730 & -30.38628 & $1.80\pm0.06 $ & $24.30\pm0.08 $ & $1.21\pm0.05 $ & & $1.81\pm0.11 $ & $24.68\pm0.14 $ & $1.32\pm0.09 $ & $7.55 _{- 0.02 }^{+ 0.04 }$ \\
4720 & 3.58980 & -30.38792 & $2.89\pm0.23 $ & $24.30\pm0.18 $ & $2.41\pm0.12 $ & & $1.39\pm0.08 $ & $24.09\pm0.13 $ & $1.77\pm0.12 $ & $7.72 _{- 0.04 }^{+ 0.04 }$ \\
4812 & 3.60840 & -30.38727 & $2.75\pm0.08 $ & $24.18\pm0.07 $ & $0.98\pm0.03 $ & & $3.12\pm0.17 $ & $24.83\pm0.13 $ & $1.14\pm0.06 $ & $8.08 _{- 0.03 }^{+ 0.02 }$ \\
5026 & 3.57398 & -30.38495 & $1.64\pm0.10 $ & $24.48\pm0.14 $ & $0.75\pm0.08 $ & & --  & -- & -- & $7.06 _{- 0.02 }^{+ 0.12 }$ \\
5444 & 3.58341 & -30.38218 & $1.83\pm0.07 $ & $24.83\pm0.09 $ & $1.03\pm0.05 $ & & $2.15\pm0.18 $ & $25.45\pm0.19 $ & $1.34\pm0.13 $ & $7.10 _{- 0.07 }^{+ 0.00 }$ \\
5740 & 3.58105 & -30.38084 & $1.58\pm0.33 $ & $25.53\pm0.47 $ & $2.18\pm0.45 $ & & $0.64\pm0.15 $ & $24.69\pm0.53 $ & $1.75\pm0.59 $ & $7.03 _{- 0.11 }^{+ 0.14 }$ \\
5869 & 3.60999 & -30.37962 & $3.42\pm0.34 $ & $24.59\pm0.23 $ & $1.97\pm0.14 $ & & $3.85\pm0.56 $ & $25.28\pm0.33 $ & $2.44\pm0.23 $ & $8.22 _{- 0.02 }^{+ 0.04 }$ \\
6641 & 3.60298 & -30.37230 & $2.03\pm0.04 $ & $23.85\pm0.05 $ & $0.98\pm0.03 $ & & $1.88\pm0.04 $ & $24.12\pm0.05 $ & $0.98\pm0.03 $ & $8.17 _{- 0.02 }^{+ 0.08 }$ \\
		\hline
	\end{tabular}
\end{table*}

We define the UDGs to have an circular effective radius ${R_{e,1.6\mu{\rm m}}>1.5}$~kpc, an effective surface brightness ${\langle\mu_{e,{1.6\mu\mathrm{m}}}\rangle>23.5}$~mag arcsec$^{-2}$, and a S\'{e}rsic index ${n_{1.6\mu{\rm m}}<2.5}$.  Figure~\ref{fig:Figure2} shows the ${\langle\mu_{e,{1.6\mu\mathrm{m}}}\rangle-R_{e,1.6\mu{\rm m}}}$ plane for 141 objects. 
Following the above criteria, we find that 22 objects are classified as UDGs. Hereafter, we refer to those 22 galaxies as UDGs and the remaining 119 galaxies as non-UDGs. Table~\ref{tab:Table1} summarizes the properties of UDGs measured in F200W. We show the spatial distribution of the UDGs in the cluster scale in Appendix~\ref{sec:AppendixA}. None of the UDGs have a spectroscopic redshift. 

{Here, to avoid excluding blue UDGs in the final sample, we do not apply color selection of the red sequence as done in the previous studies (e.g., \citealp{2017ApJ...839L..17J}; \citealp{2017ApJ...844..157L}). Nonetheless, we have confirmed that all of the 22 UDGs selected above are located on the red sequence without significant outliers, defined in the color magnitude diagram between F814W and F105W magnitudes. This indicates that the UDGs are passive galaxies which have old stellar populations.}

To inspect whether these UDGs can also be classified as UDGs in shorter wavelength images, we run {\tt GALFIT} using HST/ACS F814W imaging from the {Hubble Frontier Fields} program (29.1~mag at $5\sigma$; \citealp{2017ApJ...837...97L}) for the 22 UDGs in the same manner as F200W images. All but one object are fitted successfully, and the results are listed in Table~\ref{tab:Table1}.


\subsection{First look at UDGs in JWST NIRISS/F200W} \label{subsec:Sec4.3}

We find that 22 UDGs selected in the previous section have the median values of ${\tilde{R}_{e,1.6\mu\mathrm{m}}=2.07}$~kpc, 
${\langle\tilde\mu_{e,1.6\mu\mathrm{m}}\rangle =24.25}$~mag~arcsec$^{-2}$, and ${\tilde n_{1.6\mu{\rm m}} =1.26}$ in the F200W imaging. In the F814W imaging, the median values of structural parameters are ${\tilde{R}_{e,0.5\mu\mathrm{m}}=1.97}$~kpc, ${\langle\tilde \mu_{e,0.5\mu\mathrm{m}}\rangle =24.63}$~mag~arcsec$^{-2}$, and ${\tilde n_{0.5\mu{\rm m}} =1.33}$. In terms of the effective surface brightness, we discover that F200W images are systematically brighter than F814W images, with the median difference and the standard deviation of ${\langle\mu_{e,0.5\mu\rm{m}}\rangle-\langle\mu_{e,1.6\mu\rm{m}}\rangle=0.41\pm0.37}$ mag~arcsec$^{-2}$.   

Figure \ref{fig:Figure3} shows the comparison of two effective radii of the UDGs, measured in different filters. The majority of the UDGs have smaller radius in the F814W filter compared to the F200W filter, with the median size ratio and the standard deviation of ${R_{e,1.6\mu\mathrm{m}}/{R_{e,0.5\mu\mathrm{m}}}}=1.14\pm0.39$. We checked that this trend is unchanged even if we fix the S\'{e}rsic index to the one fitted from the F200W imaging in {\tt GALFIT}.

When we define UDGs to have $ R_{e,0.5\mu\mathrm{m}}>1.5$~kpc, and $\langle\mu_{e,0.5\mu{\mathrm{m}}}\rangle>24.0 \ (24.5)$~mag arcsec$^{-2}$ in the F814W imaging, 16 (11) out of 22 UDGs will be selected. Three UDGs are excluded because of their smaller radius in F814W imaging ($ R_{e,0.5\mu\mathrm{m}}<1.5$~kpc), as shown in Figure~\ref{fig:Figure3}. This implies that our criteria in the F200W imaging extract some UDGs that are not selected in the optical wavelengths.

\begin{figure*}
\includegraphics[width=0.98\linewidth]{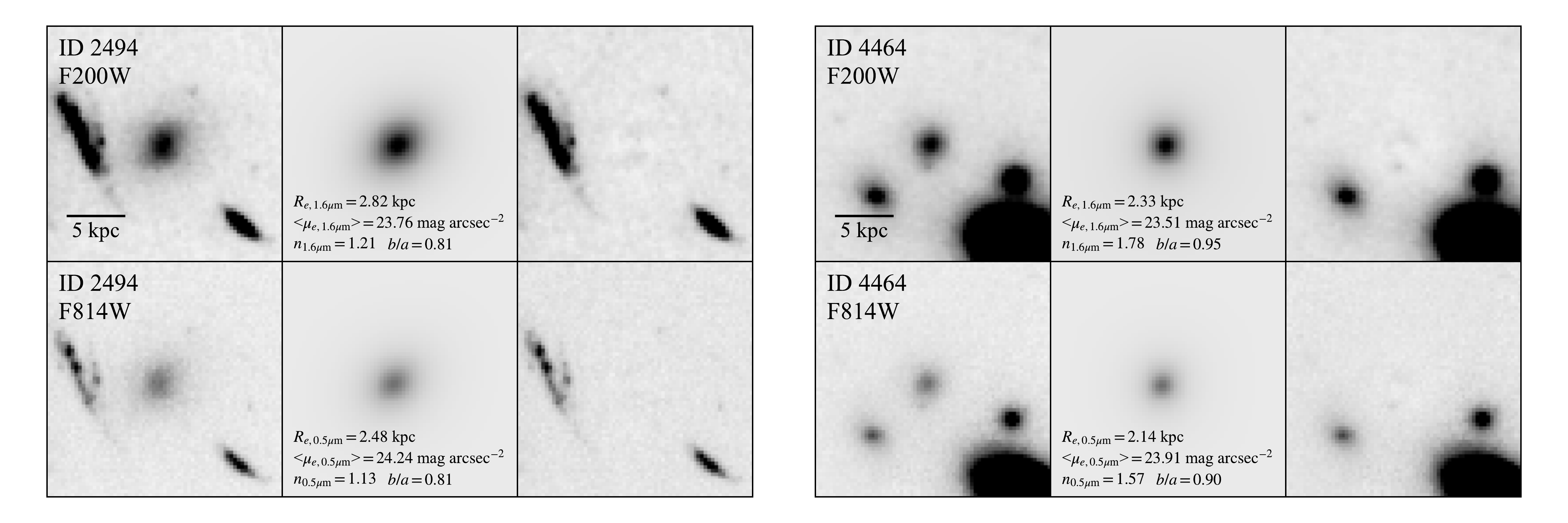}
    \caption{Two examples of the UDGs (left: ID~2494, right: ID~4464) selected from the F200W imaging. From left to right, each panel shows the data, model and residual image. For each galaxy, top and bottom panels show the images of the JWST/NIRISS F200W and HST/ACS F814W filters, respectively. Images of the F814W filters are scaled to the same zero-point magnitude of the F200W filter and shown in the same stretch.} \label{fig:Figure4}
\end{figure*}

In Figure~\ref{fig:Figure4}, we show two examples of the UDGs selected in the F200W filter, one (ID~2494) that was reported as an UDG in \cite{2017ApJ...844..157L} using the F814W filter and another (ID~4464) that was not reported in the same study, likely due to its relatively high surface brightness ($\langle\mu_{e,0.5\mu\mathrm{m}}\rangle=23.91$ mag~arcsec$^{-2}$) in the F814W filter. Both examples visually exhibit the fitted S\'{e}rsic models well reproduce the imaging data. As indicated by the best-fit structural parameters, Figure~\ref{fig:Figure4} shows that the NIR emission in the F200W image is more extended than the optical emission in F814W. A similar comparison for the rest of the 20 UDGs is presented in Figure~\ref{fig:FigureB1}.

\section{Discussion} \label{sec:Sec5}

\subsection{Comparison with the optically-selected UDGs} \label{sec:Sec5.1}

Among 27 UDGs reported in \cite{2017ApJ...844..157L}, 19 of them are within the NIRISS field. We find that 10 of their UDGs are identified as UDGs in F200W. In addition, one of their UDGs is classified as non-UDG (ID~4282) and another is A2744-DSG-z3 \citep{2023ApJ...942L...1W}.
Of the seven remaining galaxies, four are excluded since those galaxies have a photometric redshift larger than 0.5, and the other three are excluded because their structural properties do not meet the criteria in the F200W filter.
Conversely, six of our UDGs were not previously selected by \cite{2017ApJ...844..157L} using F814W.

It is worth noting that
the final list of UDGs may significantly vary due to differences in the selection process. \cite{2017ApJ...844..157L} report that they identified a smaller number of UDGs ($N=27$) compared to \citet[][$N=41$]{2017ApJ...839L..17J} despite their using the same filter for selection. The origin of discrepancy is not clear, while \cite{2017ApJ...844..157L} suggested that this is partly due to their visual inspection step. 
In this study, we have included a secondary sky component in {\tt GALFIT} which was not applied previously, and have adopted $n<2.5$ as one of the criteria while $n<4$ was used in \cite{2017ApJ...844..157L}. Therefore, we conclude that the discrepancy between the UDGs selected in this study and those reported by \cite{2017ApJ...844..157L} is likely due to a combination of the effect of looking at different wavelengths and the difference of methodologies of UDG selection.

The lack of a standard definition for what constitutes a UDG could be problematic, in the sense that less restrictive definitions may include galaxies with different origins compared to UDGs selected from restrictive definitions, and thus obscure the underlying physics that affects their formation and evolution \citep{2022ApJ...926...92V}. Suffice it to say, the UDGs that meet the criteria in both F200W and F814W filters can be regarded as a robust UDG sample. While it is intriguing to understand the morphological transition of UDGs across the wavelengths, we leave the thorough analysis and detailed discussion in future work.


\subsection{UDGs in the stellar mass-size plane} \label{subsec:Sec5.3}

\begin{figure*}
	\includegraphics[width=0.8\linewidth]{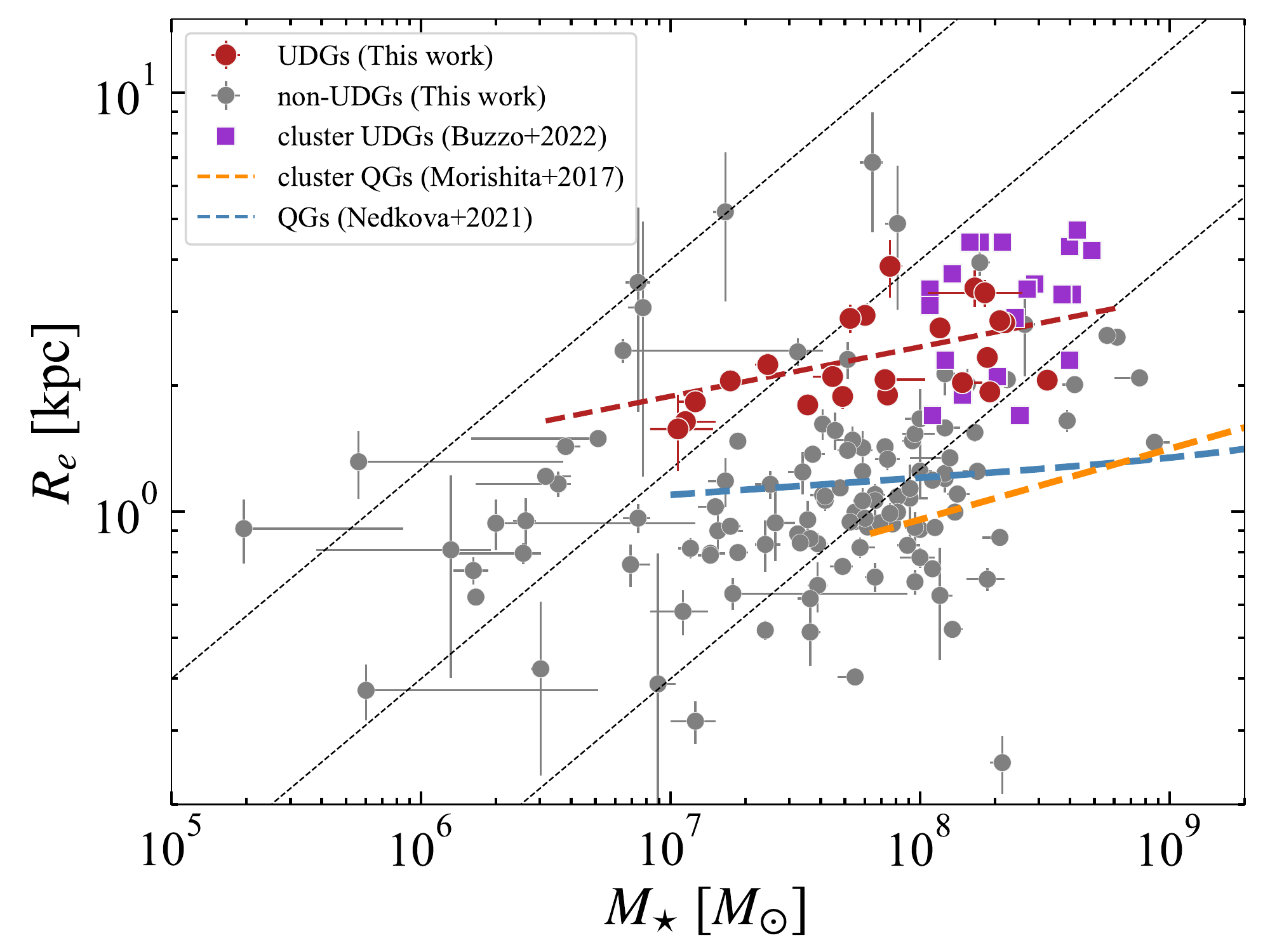}
    \caption{Distribution of the UDGs (red) and non-UDGs (grey) in Abell~2744 in the stellar mass-size plane. {$R_{e,1.6\mu\rm{m}}$ is used for the vertical axis.} {A compilation of the UDGs in nearby clusters with spectroscopic redshift} \citep{2022MNRAS.517.2231B} are shown in purple squares. The red dashed line is the result of a single power law fit for the UDGs. The orange and teal dashed lines are the best-fit law for cluster {quiescent galaxies (QGs)} in the Frontier Fields cluster ($0.2\leq z\leq0.7$; \citealp{2017ApJ...835..254M}) and QGs at $0.2\leq z\leq0.5$ \citep{2021MNRAS.506..928N}. The black dashed lines denote constant surface stellar mass density ( $\Sigma_{\star}=0.1,1,10$~$M_{\odot}\mathrm{pc^{-2}}$).} \label{fig:Figure6}
\end{figure*}

Figure \ref{fig:Figure6} shows the stellar mass-size distribution of the UDGs and non-UDGs. {For reference, the sample of cluster UDGs with spectroscopic redshift \citep{2022MNRAS.517.2231B}}, and the scaling relations of quiescent galaxies (QGs) down to the stellar mass of $M_{\star}=10^{7}M_{\odot}$ and $10^{7.8}M_{\odot}$, from \cite{2017ApJ...835..254M} and \cite{2021MNRAS.506..928N} are shown, respectively. 
Note that the sizes of UDG sample from \citet{2022MNRAS.517.2231B} and scaling relation from \cite{2021MNRAS.506..928N} are measured in optical wavelengths, while the scaling relation of \cite{2017ApJ...835..254M} is based on the HST/WFC3 F160W filter. 

The stellar masses and surface stellar mass densities of the UDGs are in the range of $10^{7}M_{\odot}\lesssim M_{\star}\lesssim10^{8.5}M_{\odot}$ and $\Sigma_{\star}\sim$ 1--10~$M_{\star}\mathrm{pc^{-2}}$, respectively. The effective radius of the F200W imaging (Table~\ref{tab:Table1}) was used for the calculation of surface stellar mass densities. We find that the range of surface stellar mass density of the UDGs are in good agreement with the definition presented in \cite{2023arXiv230304726C}. In \cite{2023arXiv230304726C}, the possible progenitors of local cluster UDGs, referred to as low surface density galaxies, are defined by the effective radius and the surface stellar mass density. The agreement implies that our UDG selection using the F200W filter is largely consistent with the physically motivated definition using the surface stellar mass density. 

The distribution of non-UDGs is consistent with the best-fit curves of low-mass QGs (\citealp{2017ApJ...835..254M}; \citealp{2021MNRAS.506..928N}), securing our size estimates. The UDGs are, by definition, located above these curves. A single power-law fitting for UDGs, using {\tt curve\_fit} ({\tt scipy}; \citealp{2020NatMe..17..261V}) reveals a correlation which is nearly flat in size with a slope of $\alpha=0.118\pm0.053$. At a given stellar mass, the size in this correlation is larger than the size in \cite{2017ApJ...835..254M} by a factor of $\sim2.6$, which reinforces that the UDGs have a distinct properties from coeval quiescent dwarf population. In terms of the flattening of the stellar-mass size distribution of QGs in the low-mass regime (e.g., \citealp{2015MNRAS.447.2603L}), the distribution of the UDGs, as well as non-UDGs, is qualitatively consistent. 


Spectroscopically confirmed UDGs in nearby clusters have stellar masses and sizes comparable to objects in Abell~2744 at the massive end. This would indicate that the spectroscopic sample of nearby UDGs is still limited to relatively massive galaxies among the entire UDG population. 


\section{Summary} \label{sec:Sec6}

We presented a search and characterization of UDGs in the central part of the Abell~2744 Frontier Field Cluster at $z=0.308$ with JWST/NIRISS F200W imaging. We define the selection criteria based on the properties measured from F200W imaging, with updated photometric and spectroscopic redshifts. 

Within the NIRISS observation field, we successfully identified 22 UDGs.
The majority of the UDGs have a brighter surface brightness (${\langle\mu_{e,0.5\mu\rm{m}}\rangle-\langle\mu_{e,1.6\mu\rm{m}}\rangle=0.41\pm0.37}$ mag~arcsec$^{-2}$) and larger size (${R_{e,1.6\mu\mathrm{m}}/{R_{e,0.5\mu\mathrm{m}}}}=1.14\pm0.39$) in F200W imaging compared to F814W imaging, which implies that the classification of UDGs is affected by not only structural parameters, but also by rest-frame wavelength at which our analysis was carried out. We find that about half of our UDG sample will not be classified as an UDG, when we adopt $ R_{e,0.5\mu\rm{m}}>1.5$~kpc and $ \langle\mu_{e,0.5\mu\rm{m}}\rangle>24.5$ mag~arcsec$^{-2}$ in the F814W imaging. The UDGs selected from the F200W filter have an estimated stellar mass range of $10^{7}M_{\odot}\lesssim M_{\star}\lesssim10^{8.5}M_{\odot}$. The distribution of the UDGs in the stellar mass-size plane is flat, and their sizes are larger than the ones in the scaling relation of coeval QGs with comparable stellar mass by a factor of $\sim2.6$. 

{This work demonstrates the detectability of UDGs at a cosmological distance by JWST, only with $\sim1/30$ of exposure time of the previous deep imaging by HST, owing to its unprecedented sensitivity in near-infrared wavelength and higher gain than HST by a factor of $(6.5\rm{m}/2.4\rm{m})^2\sim7.3$.} This remarkable facility has enabled us for a new level of investigations of distant UDG populations. Of particular interest is the spatially-resolved properties of individual UDGs. Further investigations, including deeper photometry and spectroscopy, will help us understanding the nature and the origin of UDGs.

\section*{Acknowledgements}

{We thank the anonymous referee and editor for their comments that improved the manuscript.} RI and TM would like to thank Kentaro Motohara for organizing a seminar at NAOJ in September 2022, where this work was initiated. This work is based on observations made with the NASA/ESA/CSA JWST. The data were obtained from the Mikulski Archive for Space Telescopes at the Space Telescope Science Institute, which is operated by the Association of Universities for Research in Astronomy, Inc., under NASA contract NAS 5-03127 for JWST. These observations are associated with program JWST-ERS-1324. 
BV acknowledges support from the INAF Large Grant 2022 “Extragalactic Surveys with JWST” (PI Pentericci). BM acknowledges support from Australian Government Research Training Program (RTP) Scholarships and the Jean E Laby Foundation. CG acknowledges financial support through grants PRIN-MIUR 2017WSCC32 and 2020SKSTHZ.



\section*{Data Availability}

The data are publicly available through the Mikulski Archive for Space Telescopes (MAST) portal managed by Space Telescope Science Institute.



\bibliographystyle{mnras}
\bibliography{UDG_GLASS} 



\appendix

\section{Spatial distribution of UDGs in the cluster}  \label{sec:AppendixA}

\begin{figure}
    \centering
	\includegraphics[width=0.9\columnwidth]{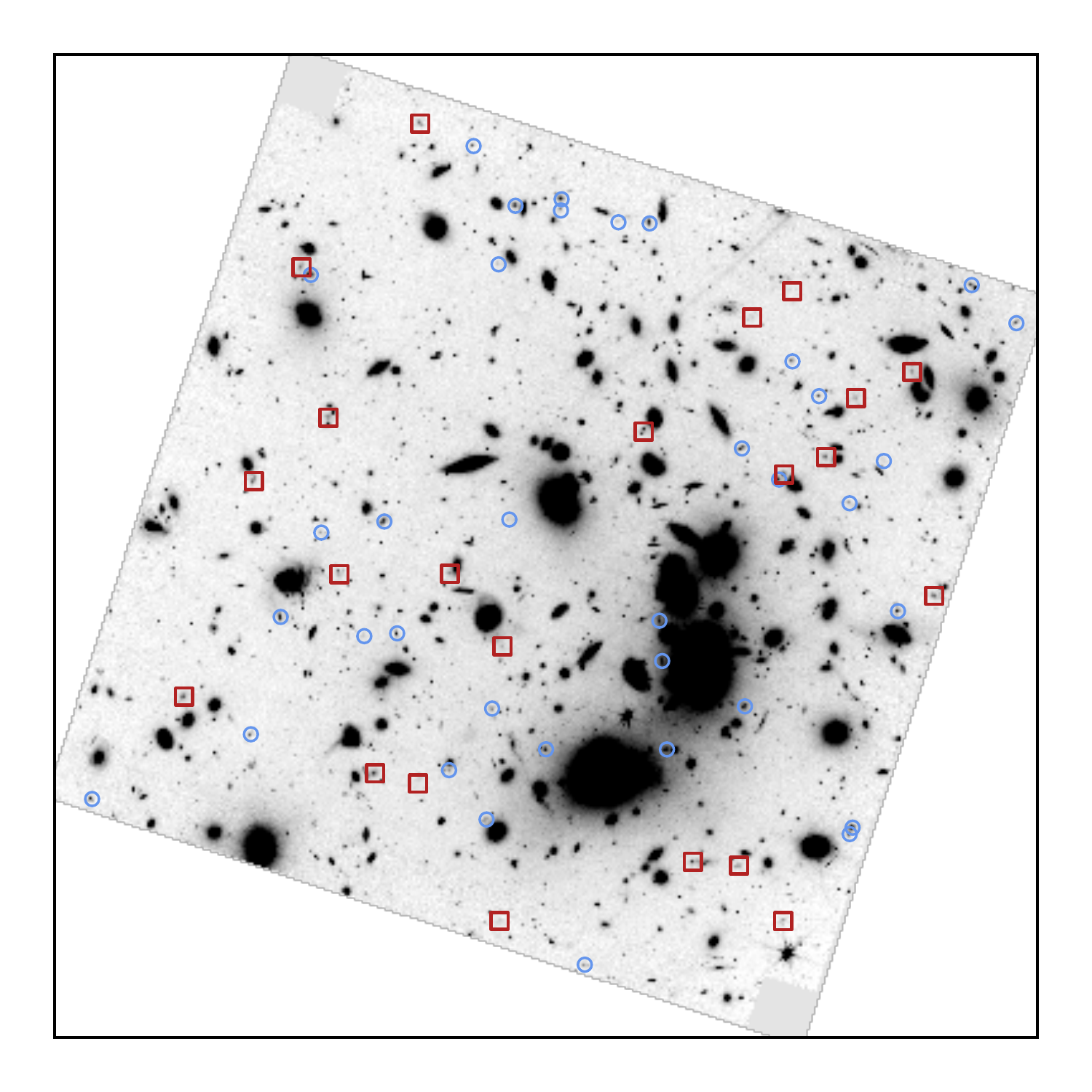}
    \caption{Spatial distribution of 22 UDGs (red squares) 
    in Abell~2744 selected by JWST/NIRISS F200W filter {($\sim0.22\times0.22$~Mpc$^{2}$ region)}. Non-UDGs with ${R_{e,1.6\mu {\rm m}}}>1.2$~kpc and $n<4$ are shown in {blue} circles. The region of $\sim0.4\times0.4$~Mpc$^{2}$ is shown.} 
    \label{fig:FigureA1}  
\end{figure}

In Figure~\ref{fig:FigureA1}, we present the overall spatial distribution of UDGs (red squares), and non-UDGs (blue circles) listed in Table~\ref{tab:TableC1}.

\section{Postage stamp images of UDGs}  \label{sec:AppendixB}

\begin{figure*}
	\includegraphics[width=0.95\linewidth]{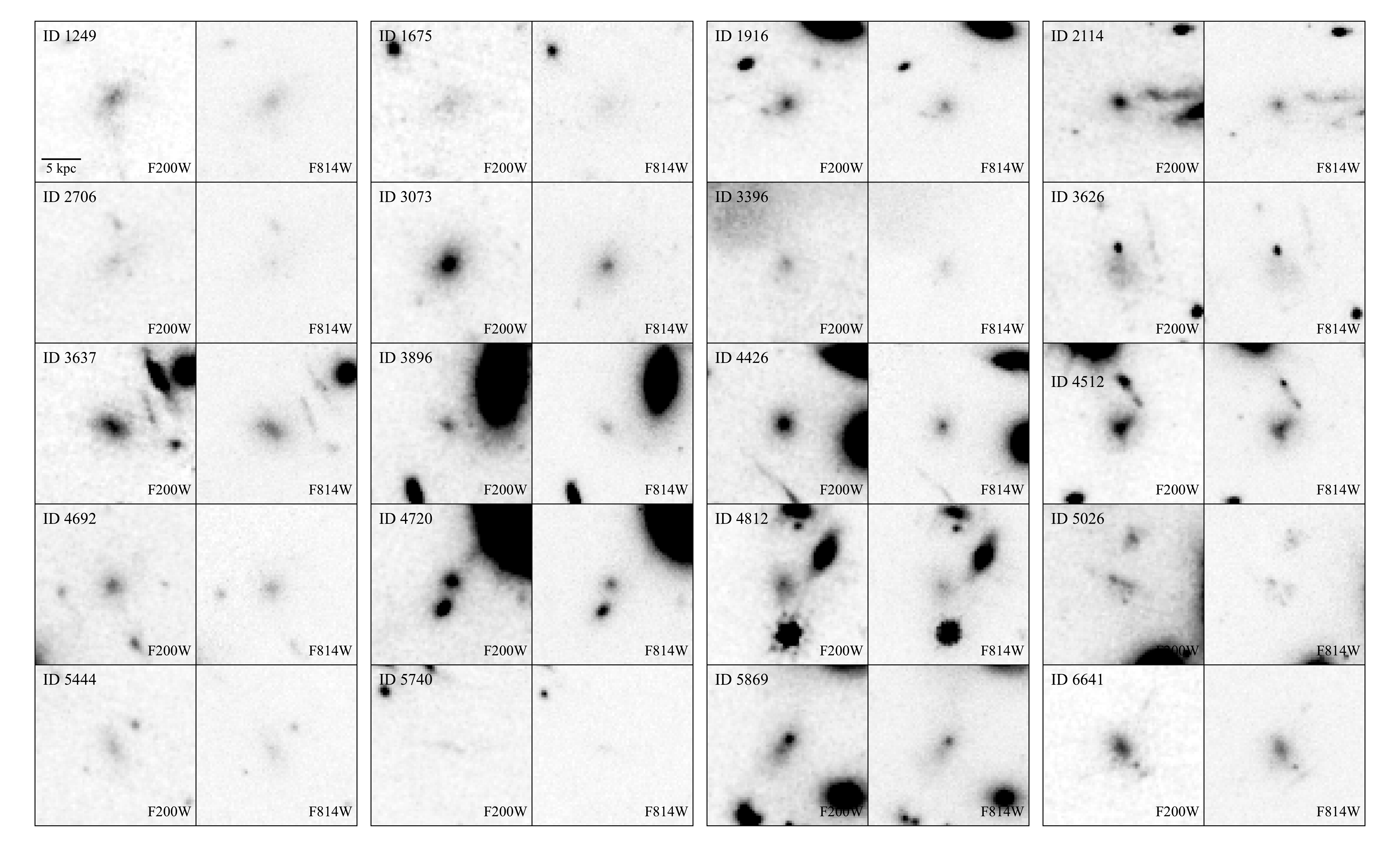}
    \caption{Postage stamp images of the UDGs selected by F200W in this study (left: F200W, right: F814W). Each panel shows a $20\times20$~kpc$^{2}$ region. Images of the F814W filters are scaled to the same zero-point magnitude of the F200W filter and shown in the same stretch.} 
    \label{fig:FigureB1}
\end{figure*}

In Figure~\ref{fig:FigureB1}, we present postage stamp images of the UDGs (except ID~2494 and ID~4464, which are shown in Figure~\ref{fig:Figure4}) in both F200W (left panel) and F814W (right panel).  


\section{The properties of non-UDGs} \label{sec:AppendixC}

{We argue that some non-UDGs may evolve into UDGs over a timescale of $\sim4$\,Gyr by expanding their size caused by tidal heating. In fact, several non-UDGs locate near the selection boundary of UDGs within the error bars in Figure~\ref{fig:Figure2}, and vice versa. Therefore, although size evolution may be insignificant \citep[e.g.,][]{2019MNRAS.490.5182L,2020MNRAS.494.1848S,2022MNRAS.514.5840P}, it is worthwhile to report these objects as possible progenitors of cluster UDGs in the nearby universe. {We select them based on a circular effective radius $ R_{e,1.6\mu\mathrm{m}}>1.2$~kpc without any selections on an effective surface brightness and a S\'{e}rsic index, since the increase of the former and decrease of the latter are expected to accompany with the size expansion.} The 1.6~\micron \ properties of these non-UDGs are summarized in Table \ref{tab:TableC1}.}

\begin{table*}
	\centering
	\fontsize{8pt}{4.3mm}\selectfont
	\caption{The properties of non-UDGs in the rest-frame 1.6~\micron \ wavelength and their stellar mass.}
	\label{tab:TableC1}
	\begin{tabular}{ccccccc}
		\hline\hline
		ID & R.A. & Decl. &${R_{e,1.6\mu {\rm m}}}$ &$\langle\mu_{e,1.6\mu{\rm m}}\rangle$ &$n_{1.6\mu{\rm m}}$ &$\log({M_{\star}/M_{\odot})}$  \\
		   & (deg) & (deg) & (kpc) & (mag~arcsec$^{-2}$) & &  \\
		\hline
		1370 & 3.59330 & -30.41510 & $1.42\pm0.14$ & $24.09\pm0.22$ & $2.04\pm0.21$ & $7.77 _{- 0.04 }^{+ 0.03 }$ \\
2206 & 3.57750 & -30.40813 & $2.61\pm0.05$ & $23.01\pm0.05$ & $1.83\pm0.03$ & $8.79 _{- 0.01 }^{+ 0.02 }$ \\
2325 & 3.57767 & -30.40846 & $1.62\pm0.14$ & $23.37\pm0.19$ & $1.80\pm0.16$ & $7.61 _{- 0.01 }^{+ 0.04 }$ \\
2427 & 3.59909 & -30.40771 & $2.31\pm0.23$ & $24.35\pm0.23$ & $2.60\pm0.19$ & $7.71 _{-0.01 }^{+0.05 }$ \\
2544 & 3.62236 & -30.40666 & $1.25\pm0.03$ & $22.74\pm0.05$ & $1.53\pm0.04$ & $8.23 _{- 0.06 }^{+ 0.02 }$ \\
2644 & 3.58385 & -30.40196 & $1.46\pm0.04$ & $22.95\pm0.06$ & $0.70\pm0.04$ & $8.94 _{- 0.02 }^{+ 0.06 }$ \\
2664 & 3.59558 & -30.40414 & $2.80\pm0.70$ & $24.00\pm0.56$ & $3.44\pm0.50$ & $8.42 _{-0.00 }^{+0.04 }$ \\
2672 & 3.58845 & -30.40415 & $1.65\pm0.10$ & $22.74\pm0.14$ & $3.00\pm0.15$ & $8.59 _{- 0.01 }^{+ 0.02 }$ \\
2686 & 3.60130 & -30.40520 & $1.47\pm0.05$ & $23.86\pm0.08$ & $1.10\pm0.05$ & $7.27 _{- 0.02 }^{+ 0.02 }$ \\
2882 & 3.61299 & -30.40337 & $2.41\pm0.18$ & $24.12\pm0.17$ & $3.57\pm0.18$ & $7.51 _{-0.02 }^{+0.04 }$ \\
3007 & 3.59876 & -30.40207 & $1.25\pm0.03$ & $23.33\pm0.05$ & $1.49\pm0.05$ & $7.77 _{- 0.01 }^{+ 0.03 }$ \\
3130 & 3.60436 & -30.39824 & $2.42\pm0.16$ & $23.96\pm0.15$ & $3.11\pm0.15$ & $6.81 _{-0.01 }^{+0.80 }$ \\
3150 & 3.58873 & -30.39965 & $1.23\pm0.12$ & $22.44\pm0.22$ & $1.70\pm0.21$ & $8.10 _{- 0.01 }^{+ 0.04 }$ \\
3458 & 3.60630 & -30.39838 & $1.49\pm0.07$ & $25.26\pm0.11$ & $0.59\pm0.07$ & $6.71 _{- 0.51 }^{+ 0.02 }$ \\
3462 & 3.58889 & -30.39760 & $1.25\pm0.16$ & $22.85\pm0.30$ & $1.74\pm0.28$ & $8.00 _{- 0.02 }^{+ 0.05 }$ \\
3524 & 3.61123 & -30.39740 & $1.48\pm0.02$ & $22.71\pm0.04$ & $1.16\pm0.03$ & $7.97 _{- 0.04 }^{+ 0.02 }$ \\
3534 & 3.57482 & -30.39712 & $1.55\pm0.08$ & $22.66\pm0.11$ & $3.71\pm0.15$ & $8.22 _{- 0.01 }^{+ 0.03 }$ \\
4024 & 3.60511 & -30.39255 & $2.63\pm0.04$ & $23.27\pm0.03$ & $1.33\pm0.02$ & $8.75 _{- 0.02 }^{+ 0.03 }$ \\
4164 & 3.60883 & -30.39311 & $1.37\pm0.05$ & $24.24\pm0.08$ & $0.95\pm0.05$ & $7.57 _{- 0.03 }^{+ 0.05 }$ \\
4282 & 3.59774 & -30.39245 & $1.48\pm0.11$ & $25.13\pm0.18$ & $0.86\pm0.10$ & $7.73 _{- 0.07 }^{+ 0.02 }$ \\
4328 & 3.57767 & -30.39163 & $1.56\pm0.16$ & $23.96\pm0.23$ & $3.71\pm0.30$ & $7.69 _{-0.02 }^{+0.04 }$ \\
4343 & 3.58183 & -30.39042 & $3.93\pm0.28$ & $24.18\pm0.16$ & $3.03\pm0.12$ & $7.66 _{-0.06 }^{+0.00 }$ \\
4380 & 3.57565 & -30.38948 & $1.24\pm0.15$ & $24.26\pm0.27$ & $2.73\pm0.34$ & $7.53 _{- 0.06 }^{+ 0.00 }$ \\
4669 & 3.58402 & -30.38884 & $1.59\pm0.02$ & $22.75\pm0.03$ & $1.36\pm0.02$ & $8.10 _{- 0.01 }^{+ 0.06 }$ \\
4963 & 3.57947 & -30.38618 & $1.43\pm0.06$ & $23.91\pm0.09$ & $1.74\pm0.08$ & $6.58 _{- 0.04 }^{+ 0.06 }$ \\
5048 & 3.58104 & -30.38441 & $1.43\pm0.02$ & $23.58\pm0.04$ & $1.00\pm0.02$ & $7.86 _{- 0.03 }^{+ 0.03 }$ \\
5378 & 3.60944 & -30.38000 & $2.09\pm0.04$ & $22.48\pm0.04$ & $2.16\pm0.03$ & $8.88 _{- 0.10 }^{+0.01 }$ \\
5502 & 3.56783 & -30.38247 & $1.34\pm0.02$ & $23.19\pm0.03$ & $0.48\pm0.02$ & $8.12 _{- 0.16 }^{+ 0.00 }$ \\
5773 & 3.57047 & -30.38053 & $1.21\pm0.05$ & $23.29\pm0.10$ & $1.47\pm0.08$ & $6.50 _{- 0.01 }^{+ 0.10 }$ \\
5838 & 3.59837 & -30.37947 & $5.20\pm2.02 $ & $26.29\pm0.89 $ & $3.88\pm0.69 $ & $7.22 _{- 0.05 }^{+ 0.02 }$ \\
6110 & 3.58946 & -30.37739 & $1.53\pm0.02$ & $22.72\pm0.03$ & $1.08\pm0.02$ & $7.98 _{- 0.05 }^{+ 0.08 }$ \\
6196 & 3.59130 & -30.37733 & $1.40\pm0.05$ & $24.46\pm0.09$ & $1.00\pm0.05$ & $7.71 _{- 0.04 }^{+ 0.10 }$ \\
6225 & 3.59737 & -30.37649 & $2.01\pm0.04$ & $23.22\pm0.04$ & $1.37\pm0.03$ & $8.62 _{- 0.05 }^{+ 0.01 }$ \\
6232 & 3.59465 & -30.37616 & $2.07\pm0.02$ & $23.08\pm0.02$ & $1.00\pm0.01$ & $8.35 _{- 0.06 }^{+ 0.01 }$ \\
6242 & 3.59469 & -30.37673 & $1.33\pm0.08$ & $23.41\pm0.13$ & $2.52\pm0.15$ & $7.87_{- 0.05 }^{+ 0.05 }$ \\
6802 & 3.59984 & -30.37345 & $1.67\pm0.30$ & $23.76\pm0.40$ & $3.57\pm0.49$ & $8.17 _{-0.02 }^{+0.08 }$ \\
    \hline
	\end{tabular}
\end{table*}


\bsp	
\label{lastpage}
\end{document}